\documentclass[a4paper]{jpconf}
\usepackage{graphicx}
\def \be{\begin{equation}}
\def \ee{\end{equation}}
\def \bes{\begin{eqnarray}}
\def \ees{\end{eqnarray}}
\begin{document}
\title{Quantum gravity as a group field theory: a sketch}

\author{Daniele Oriti}

\address{Department of Applied Mathematics and Theoretical Physics \\
  Centre for Mathematical Sciences, University of Cambridge \\
  Wilberforce Road, Cambridge CB3 0WA, England, EU}

\ead{d.oriti@damtp.cam.ac.uk}

\begin{abstract}
We give a very brief introduction to the group field theory approach
to quantum gravity, a generalisation of matrix models for
2-dimensional quantum gravity to higher dimension, that has emerged
recently from research in spin foam models.
\end{abstract}

\section{Introduction: sum-over histories quantum
  gravity and the 3rd quantization idea}

Quantum gravity remains the elusive dream of fundamental theoretical
physics. The multitude of approaches that are currently
pursued is vast \cite{book}. Some of these approaches attempt to realise on solid
grounds the idea of defining quantum gravity as a sum-over-histories of
the gravitational field. This would work roughly
as follows.
Consider a compact 4-manifold (spacetime) with trivial topology
$\mathcal{M}$ and all the possible geometries (spacetime metrics up to
diffeomorphisms) that are compatible with it. The partition function of the theory would then be defined \cite{hartle} by an integral over all possible 4-geometries, with a
diffeomorphism invariant measure, weighted by the exponential of the
action of General Relativity. For computing quantum gravity transition
amplitudes, one would
instead consider a manifold $\mathcal{M}$, again of trivial topology, with two disjoint boundary
components $S$ and $S'$ and given boundary data, i.e. 3-geometries, on
them: $h(S')$ and $h'(S')$, and define the transition amplitude by:
\be
Z_{QG}\left(h(S),h'(S')\right)=\int_{g(\mathcal{M}\mid h(S),h'(S'))}\mathcal{D}g \,e^{i\,S_{GR}(g,\mathcal{M})}
\ee
i.e. by summing over all 4-geometries inducing the given 3-geometries
on the boundary.
The expression above is purely formal, in absence of a
rigorous definition of a suitable measure in the space of
4-geometries; also, its physical
interpretation is challenging,
given that the formalism seems to be bound to a cosmological setting, where our usual interpretations of quantum
mechanics are not applicable. This has not prevented
physicists to propose generalisations. 
Why not to include also spacetime topology into the set of dynamical
variables and allow for spatial topology changing
processes? One could just extend the sum over geometries
above to include a sum over different
manifolds, but faces the impossibility of classifying topologies
in 4 dimensions, and no clearcut
criterion could be found for assigning a weight  to each topology in the sum. A \lq\lq 3rd
quantization\rq\rq formalism \cite{strogidd,guigan} was then proposed, in
which the topology changing processes are described as a
field theoretic \lq interaction of universes\rq. The idea is to define
a (scalar) field $\phi(^3h)$ in superspace $\mathcal{H}$, i.e. in the space of all possible
3-geometries (3-metrics $^3h_{ij}$ up to diffeos), with action:
\be
S(\phi)=\int_{\mathcal{H}}\mathcal{D} ^3h\,\phi(^3h)\Delta\phi(^3h) + \lambda \int_{\mathcal{H}}\mathcal{D} ^3h\,\mathcal{V}\left(\phi(^3h)\right)
\ee
with $\Delta$ being the Wheeler-DeWitt operator of canonical
gravity here defining the free propagation of the
field, while $\mathcal{V}(\phi)$ is a generic, e.g. cubic,
and possibly non-local (in superspace) interaction term, governing topology change. The partition function $Z=\int\mathcal{D} \phi 
e^{-S(\phi)}$, produces, in perturbative expansion in Feynman graphs, the
quantum gravity path integral for trivial topology, representing a sort
of one particle propagator, thus a Green function for the
Wheeler-DeWitt equation, plus a sum over topologies with definite weights. Note two
features of this formalism: 1) the
classical field equations will be a
non-linear extension of the Wheeler-DeWitt equation of canonical
gravity, due to the interaction term in the action, i.e. due to topology change; 2) the perturbative 3rd quantized vacuum
of the theory will be the \lq\lq no spacetime\rq\rq state, and not any
state with a semiclassical smooth geometric interpretation, e.g. Minkowski space.

\section{Modern approaches: matrix
  models, dynamical triangulations, spin foams}
These \lq\lq 3rd
quantization\rq\rq ideas were realised rigorously, although in a much
simpler context, in matrix models for 2-d Riemannian quantum
gravity \cite{matrix}. Consider the action
\be
S(M)=\frac{1}{2}\tr{M^2}\,-\,\frac{\lambda}{3!\sqrt{N}}\,\tr{M^3}
\ee
for an $N\times N$ hermitian matrix $M_{ij}$, and the associated
partition function $Z=\int dM e^{-S(M)}$. This can be expanded in Feynman diagrams; propagators and vertices
of the theory can be represented diagrammatically, and Feynman diagrams,
obtained as usual by gluing vertices with propagators, are given by
{\it fat graphs} of {\bf all} topologies. Moreover, propagators and vertices can be
understood as topologically dual to edges and triangles of a
2-dimensional simplicial complex dual to the whole fat graph; one can
then define 2d quantum gravity, via the
perturbative expansion for the matrix model above, as sum over {\bf
  all 2d triangulations} $T$ of {\bf all topologies}. 
\begin{figure}[here]
\vspace{-0.8cm}
\setlength{\unitlength}{1cm}
\begin{minipage}[b]{7cm}
\includegraphics[width=6cm, height=2cm]{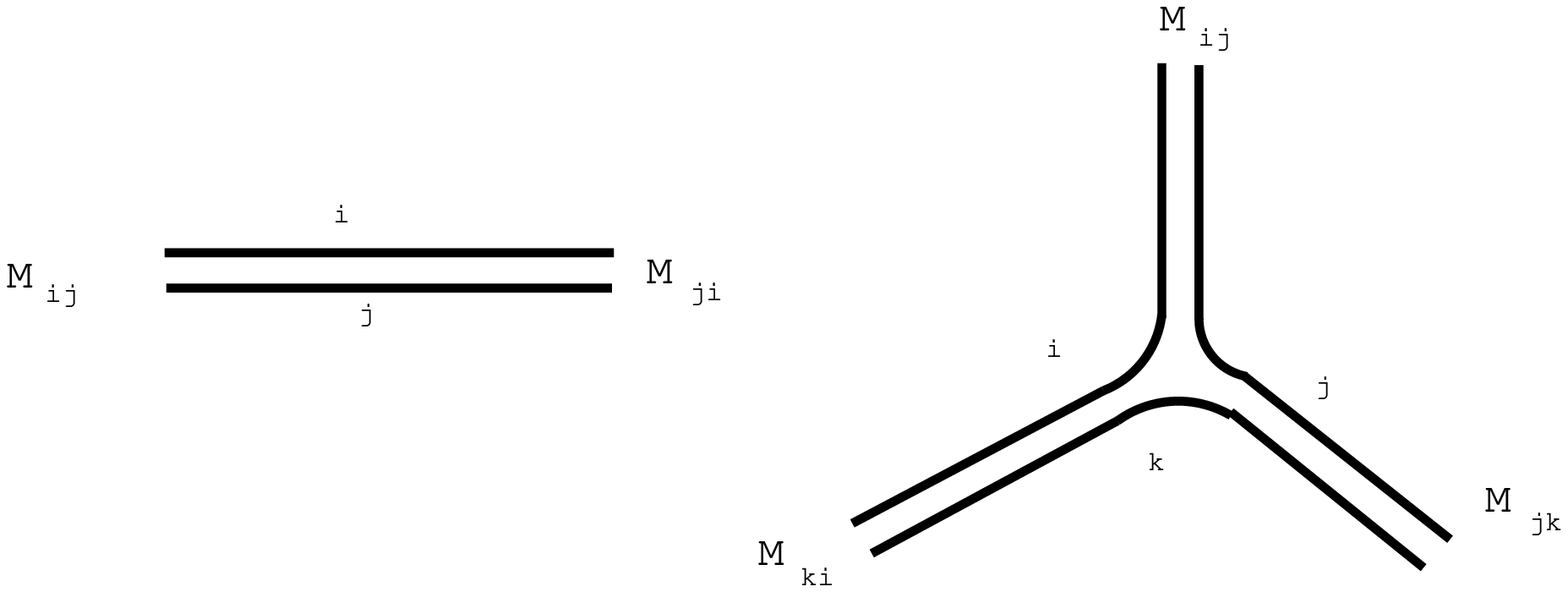}
\caption{\label{fig:matrixpropvertex} Propagator and vertex}
\end{minipage}
\hfill
\begin{minipage}[b]{7cm}
\includegraphics[width=6cm, height=2cm]{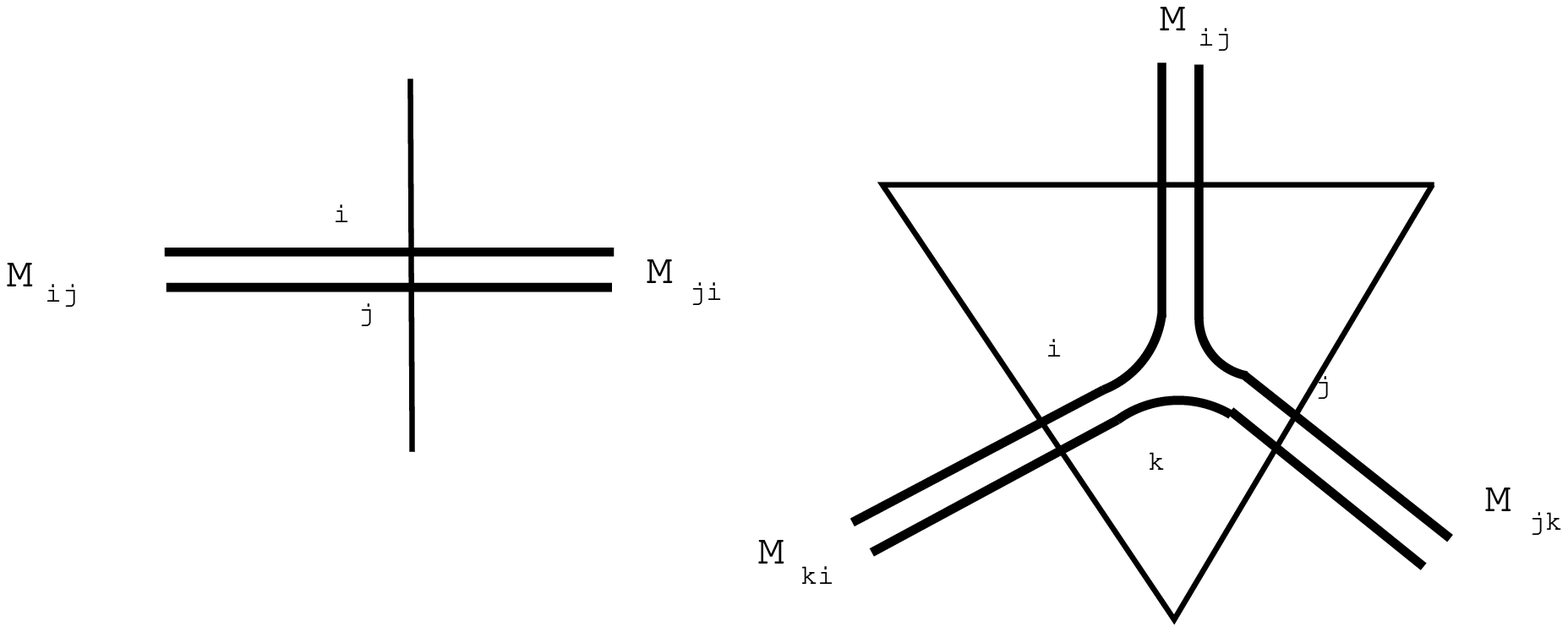}
\caption{\label{fig:matrixdualpropvertex} Dual picture}
\end{minipage}
\end{figure}

\vspace{-0.5cm}
Each Feynman diagram amplitude can be related to
the Regge action for simplicial gravity for fixed
edge lengths $N$ and positive cosmological constant, and the partition function is:
\be
Z=\int dM e^{-S(M)}=\sum_{T} \frac{1}{sym(T)}\lambda^{n_2(T)}N^{\chi(T)}
\ee   
where $sym(T)$ is the order of symmetries of the triangulation $T$, $n_2$ is the number of triangles in it, and $\chi$ is the Euler characteristic of the same triangulation.
Many results have been obtained over the years for this class of
models, for which we refer to the literature \cite{matrix}, among
which the link with continuum formulations of 2d quantum
gravity. Matrix models manage to treat
topology as a dynamical variable in a simplicial context, while rigorously defining a simplicial path integral formulation of
quantum gravity for given topology. This raised the hopes that similar
techniques and structures could be used to define a path integral for
gravity also in higher dimensions and possibly for the Lorentzian
signature. The dynamical triangulations approach \cite{DT} is defined
exactly on these bases. A path integral for gravity
(for fixed topology) can be given meaning in a simplicial setting,
modelling D-dimensional spacetime as the simplicial complex with fixed edge length $a$, thus
encoding the degrees of freedom of the gravitational field in the
combinatorics of the simplicial complex only, and defining the
partition function as a sum over all triangulations with fixed
topology weighted by the Regge action for
gravity:
\be
Z(G,\lambda,a)=\sum_{T}\frac{1}{sym(T)}e^{iS_R(T,G,\Lambda,a))}
\ee
where $G$ is the gravitational constant and $\Lambda$ is a
cosmological constant. In the Lorentzian case one also distinguishes
between spacelike and timelike edges, and imposes some additional restrictions on
the topology considered and on the way the triangulations are
constructed. This leads to a well-defined partition function of gravity,
that can be dealt with both analytically and numerically to extract physical predictions. In particular, one may
look for a continuum limit of the theory, corresponding to the limit
$a\rightarrow 0$ accompanied by a suitable renormalisation of the
constants of the theory $\Lambda$ and $G$.
And exciting recent results \cite{DT} seem
to indicate that, in the Lorentzian context and for trivial topology,
a smooth phase with the correct dimensionality is obtained even in 4
dimensions, which increases the confidence in the correctness of the strategy
adopted. 
Spin foam models \cite{review, alex} are yet another implementation
of the path integral idea. Here spacetime is
represented by a 2-complex (a collection of vertices,
edges joining them and faces bounded by these edges), the histories of
the gravitational field (4-geometries) are given by {\bf spin foams}, i.e. by these 2-complexes labelled
with irreps $\rho$ of the Lorentz group assigned
to their faces, and boundary data (3-geometries) are {\bf spin networks}, i.e. graphs (boundary
of the 2-complexes) labelled by irreps of the same type, assigned to the links of the
graph. 
\begin{figure}[here] 
\setlength{\unitlength}{1cm}
\vspace{-0.4cm}
\begin{minipage}[b]{5cm}
\includegraphics[width=4cm, height=3cm]{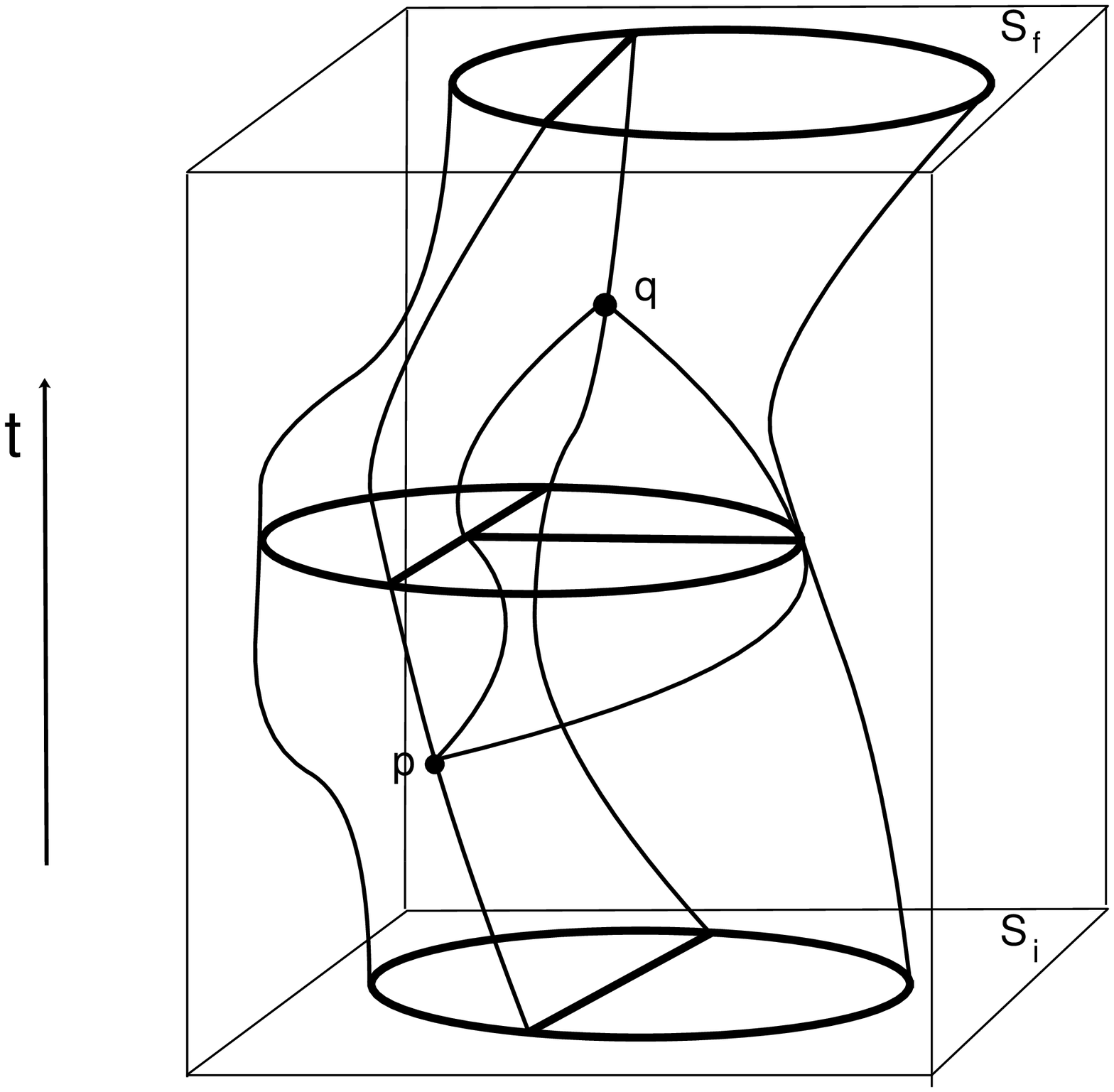}
\caption{\label{fig:spinfoam} A spin foam}
\end{minipage}
\hfill
\begin{minipage}[b]{6cm}
\includegraphics[width=3.5cm, height=2.5cm]{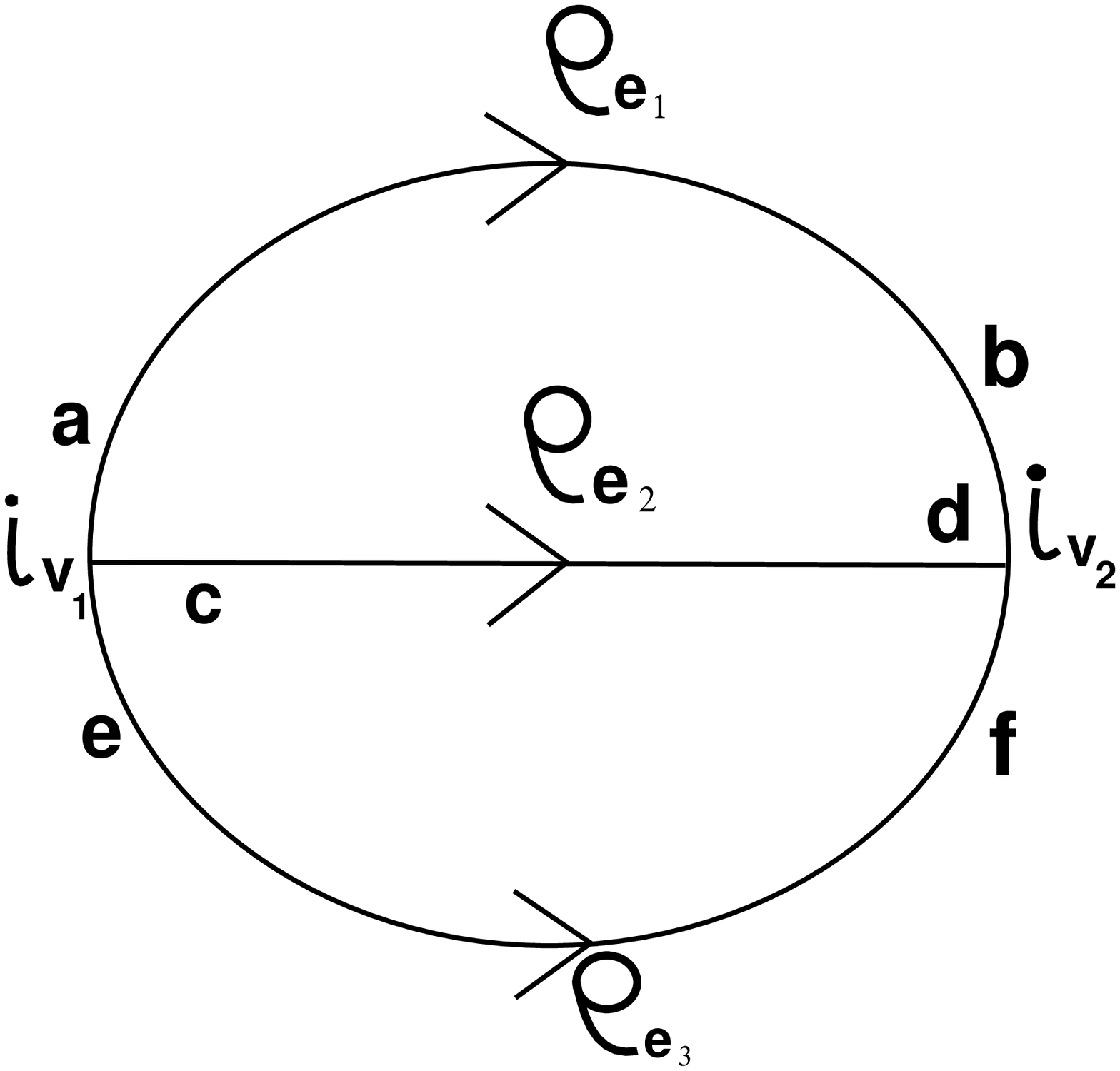}
\caption{A spin network}
\end{minipage}
\end{figure}

\vspace{-0.3cm}
The geometric degrees of freedom are thus encoded in
purely combinatorial and algebraic data, and the model is defined
by an assignment of a quantum probability amplitude (here factorised
in terms of face, edge, and vertex contributions) to each spin foam $\sigma$, and by a sum over both 2-complexes and representations, for given boundary spin networks $\Psi,\Psi'$:
$$
Z=\sum_{\sigma\mid\Psi,\Psi'}w(\sigma)\sum_{\{\rho\}}\prod_{f}A_{f}(\rho_f)\prod_{e}A_{e}(\rho_{f\mid
  e})\prod_{v}A_{v}(\rho_{f\mid v}). $$
The crucial point is how to choose the quantum amplitudes, e.g. from some
discretization of a classical action for gravity. Whatever the starting point, one would have a
rigorous implementation of a sum-over-histories for gravity in a
combinatorial-algebraic context, and should then prove that one can
both analyse fully the quantum domain, and recover classical and
semi-classical results in some appropriate approximation. 
Spin foam models have grown to a promising approach to quantum
gravity only recently, but a multitude of results have been already
obtained in this context, for which we refer to the literature \cite{review, alex}. 

\section{The group field theory formalism}
Let us now discuss how the dream of third quantization is realised (at
least tentatively) in group field theories, by extending to higher
dimensions the structures of matrix models.
 
\subsection{General structure of GFTs}
The general structure of a group field theory, independently of the
spacetime signature, is as follows
\cite{laurentgft,iogft,review,alex}. Consider
a (real or complex) scalar field over D copies of a group manifold $G$ (for quantum gravity, the Lorentz group) whose classical dynamics is
governed by the action: 
\begin{eqnarray*} S_D(\phi, \lambda)= \frac{1}{2}\prod_{i=1,..,D}\int
  dg_id\tilde{g}_i
  \phi(g_i)\mathcal{K}(g_i\tilde{g}_i^{-1})\phi(\tilde{g}_i) +
  \frac{\lambda}{(D+1)!}\prod_{i\neq j =1}^{D+1}
  \phi(g_{1j})...\phi(g_{D+1 j})\,\mathcal{V}(g_{ij}g_{ji}^{-1}),
\end{eqnarray*}
where of course the exact choice of the kinetic and interaction
operator is what defines the model. One usually imposes on the field invariance under simultaneous multiplication by a group element, and under
their permutations (maybe only even ones).  
The quantum theory is coded in the partition function,
defined again by its perturbative expansion in Feynman graphs:
$$ Z\,=\,\int
\mathcal{D}\phi\,e^{-S[\phi]}\,=\,\sum_{\Gamma}\,\frac{\lambda^N}{sym[\Gamma]}\,Z(\Gamma).
$$
As in ordinary QFT, the field can be
expanded in modes (momenta), and the Feynman amplitudes written in both configuration and momentum space; the modes of the
field are labelled by representations of the group, whose elements
define the configuration space of the field. As in matrix models
Feynman graphs are represented by {\it fat
  graphs} given by $D$ parallel lines for each propagators being
re-routed at each vertex of interaction, and  again one can give a dual interpretation of
propagators and vertices in terms of (D-1)-simplices and D-simplices
respectively. In this way, the Feynman graphs are cellular complexes
(links identifying faces, that in turn close to form 2-cells, etc)
that are topologically dual to D-dimensional triangulated (pseudo-)manifolds
of {\bf all topologies}. The Feynman amplitudes of the theory turn out
to be given, when all fields are expanded in
representations of the group $G$ and thus the amplitude are
given as a sum over these representations of appropriate functions of
them, by spin foam models, with
representation data assigned to the
faces of the Feynman graph. 
When one restricts the sum over Feynman graphs to {\it tree level},
only manifolds with trivial topology are included \cite{laurentgft},
then boundary data and transition amplitudes acquire a canonical
interpretation: boundary data define canonical quantum states of
gravity and the transition amplitude between them defines a {\it
  projection} onto physical states, i.e. those satisfying the
Hamiltonian constraint of canonical gravity, and thus the
inner product of the canonical theory.
The observables of the theory are gauge
invariant (with respect to the symmetries of the action) functions of the field operators; for example, polynomial
functionals can be expanded in spin networks. In particular, one
defines transition amplitudes by inserting appropriately
contracted field operators as observables in the partition function,
as customary in field theory, and this produces (after perturbative
expansion, and in momentum space) a sum over spin foams with spin
network states on the boundary, reflecting the combinatorics of field
operators in the observables whose expectation value is being evaluated. 
All this has a consistent quantum {\bf geometric interpretation}: each field
is understood as a 2nd quantized (D-1)-simplex, with its $D$ arguments
representing the (D-2)-simplices on its boundary; the evolution and
interaction of these fundamental building blocks (quanta of space),
that can in turn be phrased in terms of their creation/annihilation,
and represented diagrammatically in Feynman graphs,
is what generates a D-dimensional spacetime; depending on the actual
graph considered (a possible spacetime history of interactions of the quanta of
space), the resulting spacetime can have arbitrary topology and
complexity, depending on the complexity of the states
involved. The representations labelling the Feynman
graphs and being summed over in the partition function are also
interpreted geometrically: they
represent the volume of the (D-2)-simplices they correspond to, while
the group elements integrated over in configuration space correspond
to holonomies of the gravity connection. In
addition, the amplitude for each process, i.e. for each discrete
spacetime, can be related to a discretization of the gravity action on
that specific spacetime. 
\subsection{An example: 3d Riemannian Quantum Gravity}
An explicit realisation of the formalism will make clear the above
picture. We consider explicitely the 3d Riemannian quantum
gravity case (where the local gauge group is $SU(2)$), whose group field theory formulation was first given by
Boulatov \cite{boulatov}. The other existing models in 3 and 4
dimensions have a very similar formulation \cite{DP-F-K-R,PR}.
Consider the real field: $\phi(g_1,g_2,g_3): (SU(2))^{3} \rightarrow
\mathcal{R}$, with the symmetry: $\phi(g_1 g, g_2 g, g_3 g) =
\phi(g_1,g_2,g_3)$, imposed through the projector: $\phi(g_1,g_2,g_3)
= P_g \phi(g_1,g_2,g_3) = \int dg \,\phi(g_1 g,g_2 g,g_3 g)$ and the
symmetry: $\phi(g_1,g_2,g_3) =
\phi(g_{\pi(1)},g_{\pi(2)},g_{\pi(3)})$, with $\pi$ an arbitrary
permutation of its arguments. In this specific
case, the interpretation is that of a 2nd quantized triangle with its
3 edges corresponding to the 3 arguments of the field; the irreps of
$SU(2)$ labelling these edges in the mode expansion of the field have
the interpretations of edge lengths. The classical
theory is defined by the action:
\begin{eqnarray*} S[\phi] &=& \frac{1}{2}\int
dg_1..dg_3 [P_g\phi(g_1,g_2,g_3)]^2 \,- \\ &-&\frac{\lambda}{4!}\int
dg_1..dg_6 [P_{h_1}\phi(g_1,g_2,g_3)][P_{h_2}\phi(g_3,g_5,g_4)]
[P_{h_3}\phi(g_4,g_2,g_6)][P_{h_4}\phi(g_6,g_5,g_1)], \end{eqnarray*}
whose structure is chosen so to reflect the combinatorics of a 3d
triangulations, with four triangles (fields) glued along their edges
(arguments of the field) pairwise, to form a tetrahedron (vertex term),
and two tetrahedra being glued alog their common triangles (kinetic term).
The partition function is defined in terms of perturbative expansion in Feynman graphs:
$$Z\,=\,\int d\phi\,e^{-S[\phi]}\,=\,\sum_{\Gamma}\,\frac{\lambda^N}{sym[\Gamma]}\,Z(\Gamma). $$
In order to construct explicitely the quantum amplitudes for the Feynman graphs, we need to identify propagator and vertex amplitude. These are read
out from the action to be:
\begin{eqnarray*} &\mathcal{P}& = \mathcal{K}^{-1} =
    \mathcal{K} = \sum_{\pi}\int dg d\bar{g} \,\,\delta(g_1 g
  \bar{g}^{-1} \tilde{g}_{\pi(1)}^{-1})\delta(g_2 g \bar{g}^{-1}
  \tilde{g}_{\pi(2)}^{-1})\delta(g_3 g \bar{g}^{-1}
  \tilde{g}_{\pi(3)}^{-1}), \\ 
&\mathcal{V}& = \int dh_i \,\delta(g_1 h_1
  h_3^{-1} \tilde{g}_1^{-1}) \delta(g_2 h_1 h_4^{-1} \tilde{g}_2^{-1})
  \delta(g_3 h_1 h_2^{-1} \tilde{g}_3^{-1}) \delta(g_4 h_2
  h_4^{-1} \tilde{g}_4^{-1}) \delta(g_5 h_2 h_3^{-1} \tilde{g}_5^{-1})
  \delta(g_6 h_3 h_4^{-1} \tilde{g}_6^{-1})
  \;\;\;\;\;\;\;\;\;\;\;\;\end{eqnarray*}
See the picture for a diagrammatic representation, with boxes
  representing the integration over the group.
\begin{figure}[here]
\setlength{\unitlength}{1cm}
\vspace{-1.7cm}
\begin{minipage}[t]{6cm}
\begin{picture}(5.0,4.5)
\includegraphics[width=4.3cm, height=3.2cm]{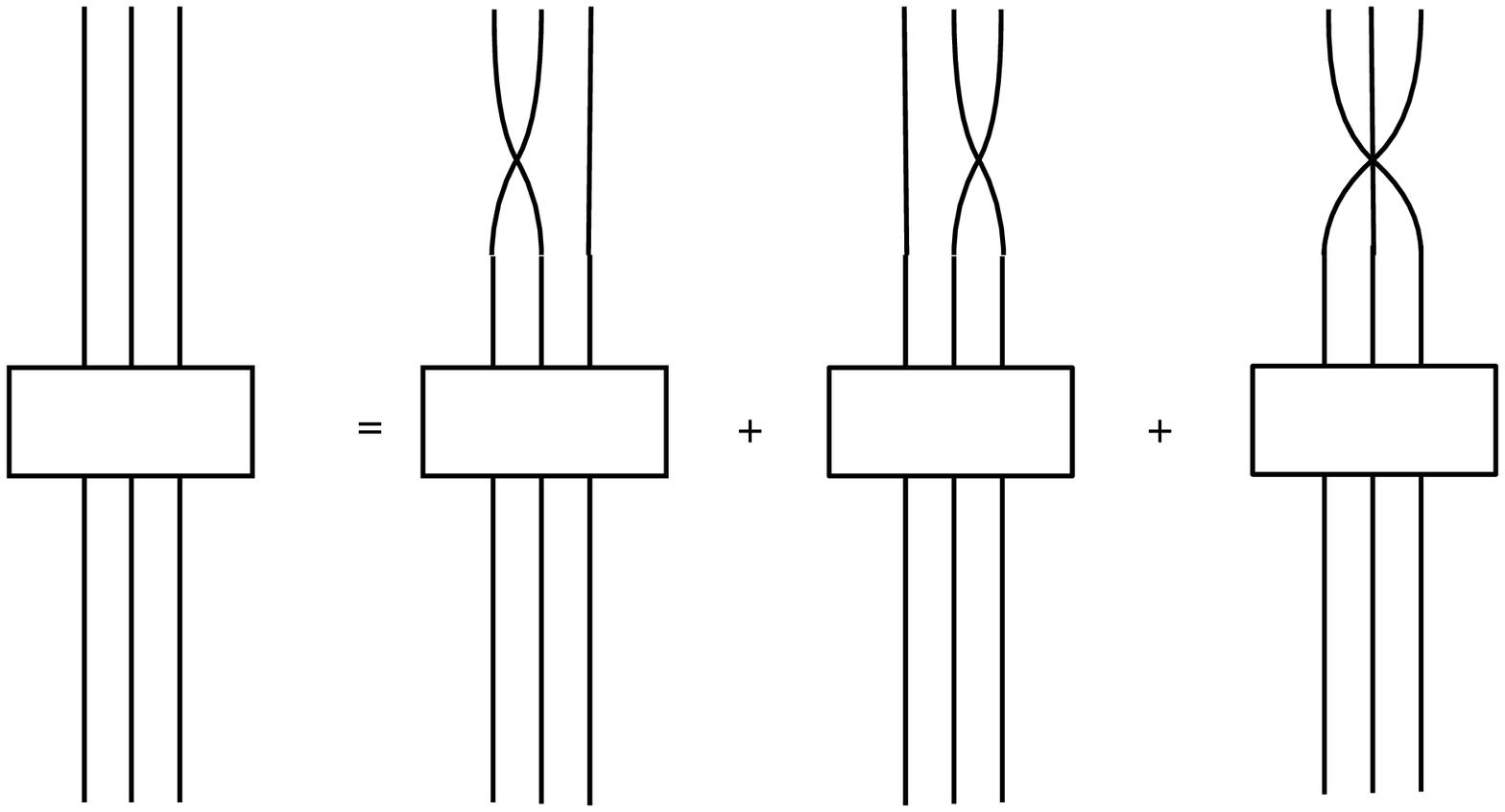}
\end{picture} \par
\caption{Propagator}
\end{minipage}
\hfill
\begin{minipage}[t]{6cm}
\begin{picture}(3.0,4.5)
\includegraphics[width=3cm, height=3cm]{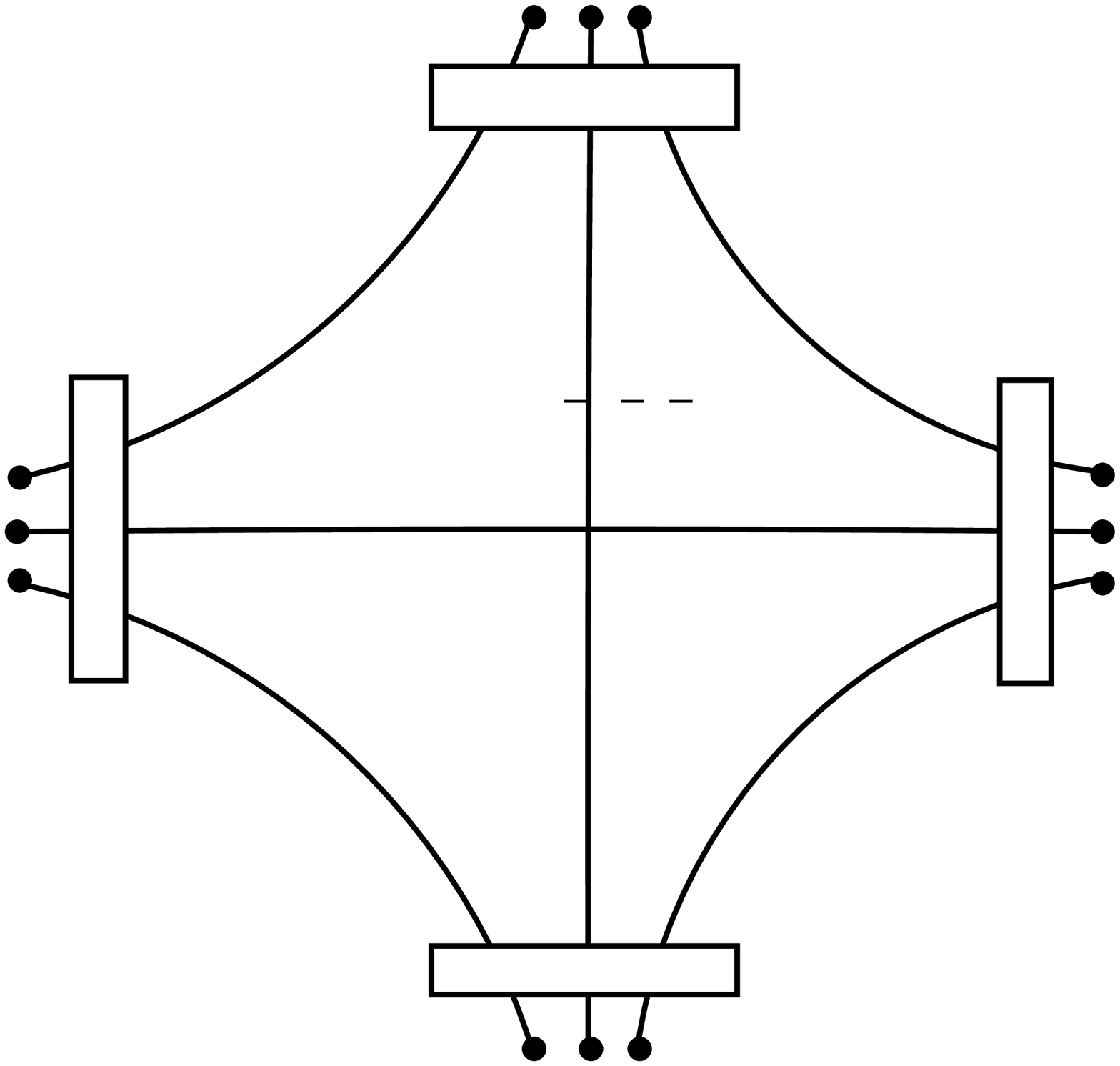}
\end{picture} \par
\caption{Interaction vertex}
\end{minipage}
\end{figure}
\vspace{-0.4cm}
The Feynman graphs are obtained as usual by gluing vertices with
propagators. Let us see how they look like. Each line in a propagator
goes through several vertices and for closed graphs it comes back to the original point, thus identifying a 2-cell; these 2-cells, together
with the set of lines running parallel in each propagator, and the set
of vertics of the graph, identify a 2-complex for each given Feynman
graph. Each of these 2-compelxes is dual to a 3d triangulation, with
each vertex correspondings to a tetrahedron, each link to a triangle
and each 2-cell to an edge of the triangulation (see picture). 
The {\bf sum over Feynman graphs} is thus equivalent to a {\bf sum
over 3d triangulations of any topology}.
\begin{figure}[here]
\setlength{\unitlength}{1cm}
\vspace{-1.2cm}
\begin{minipage}[t]{5cm}
\includegraphics[width=2.5cm, height=2.5cm]{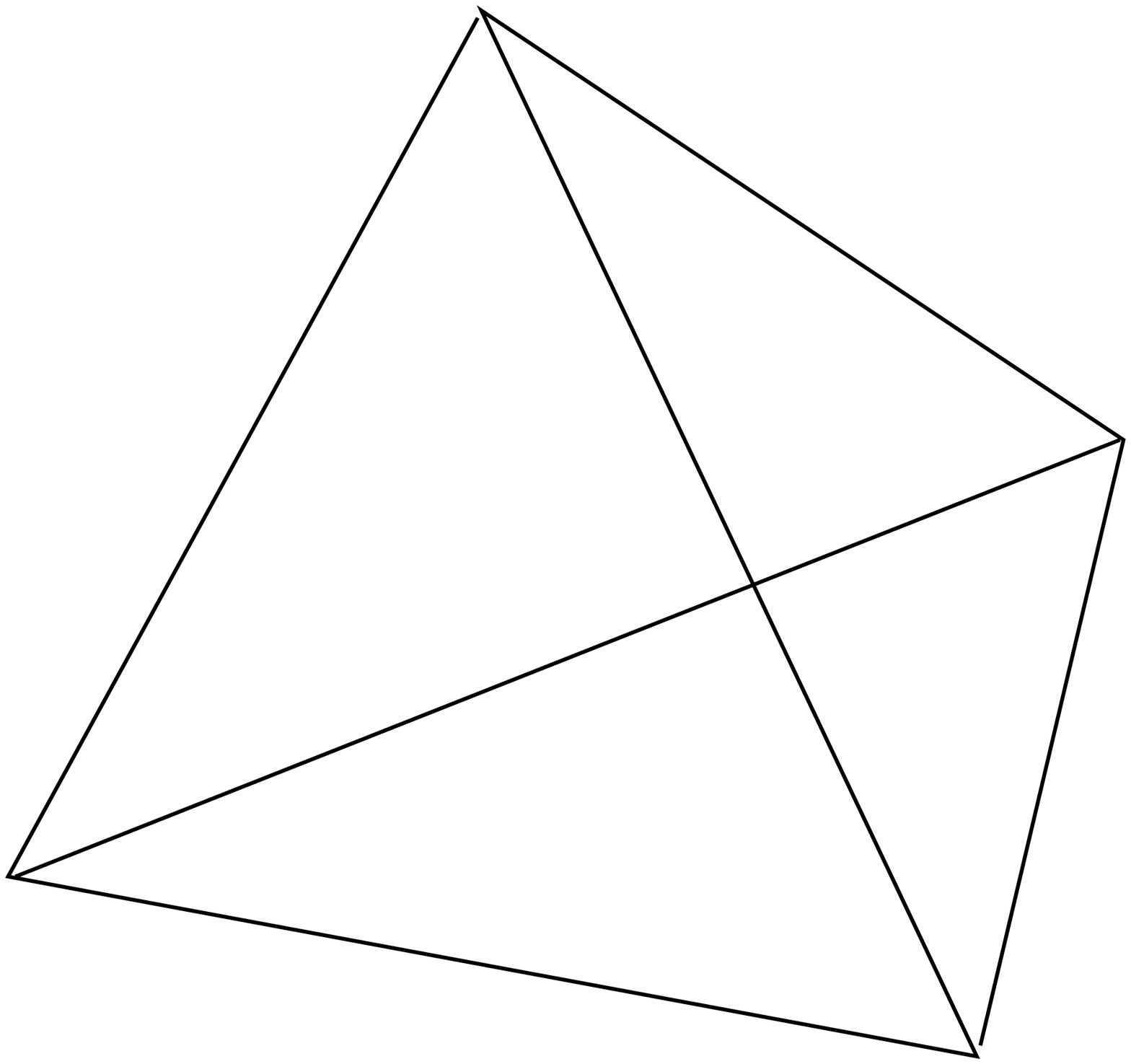}
\caption{Tetrahedron}
\end{minipage}
\hfill
\begin{minipage}[t]{5cm}
\includegraphics[width=2.5cm, height=2.5cm]{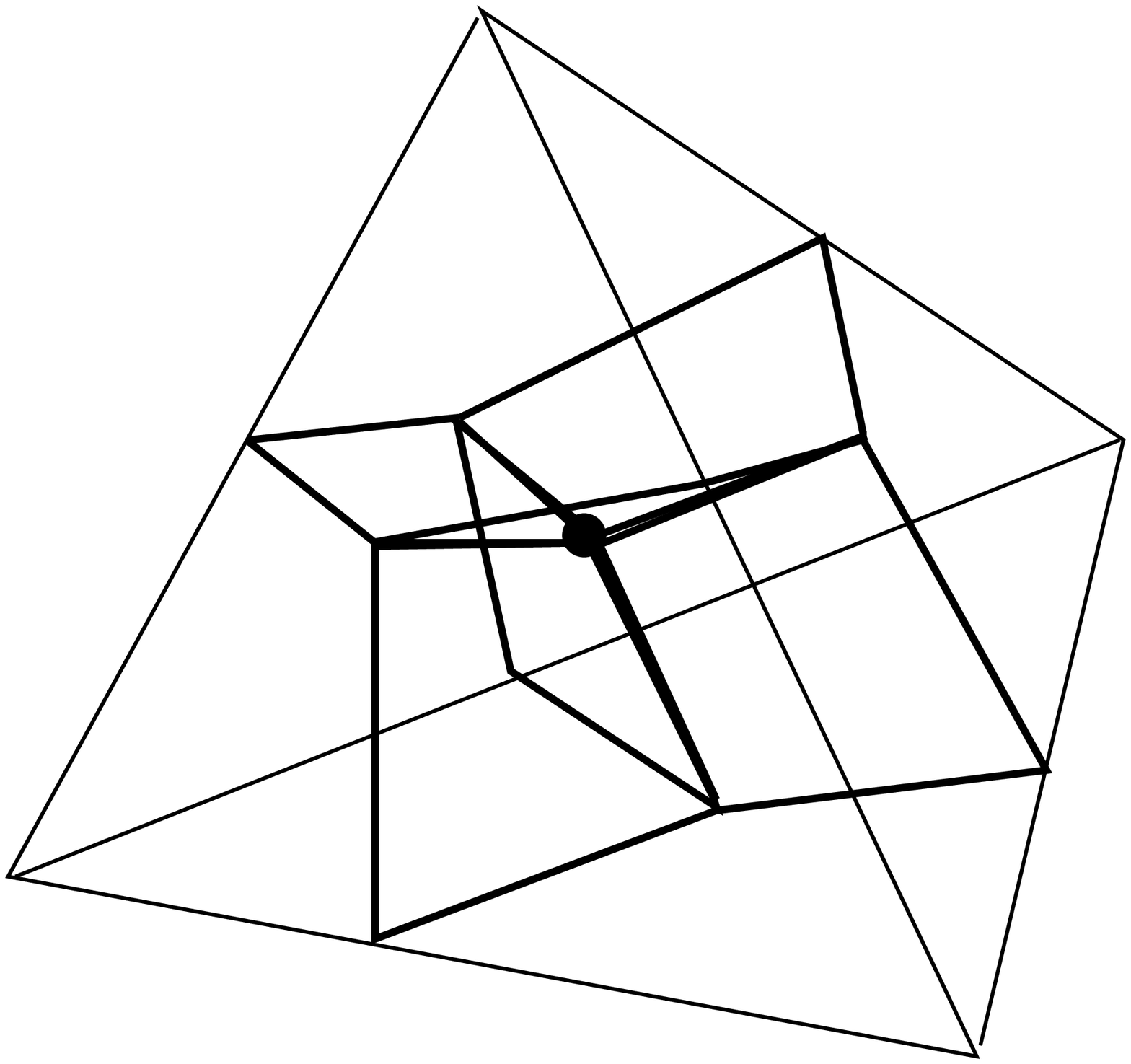}
\caption{Tetrahedron + dual 2-complex}
\end{minipage}
\hfill
\begin{minipage}[t]{5cm}
\includegraphics[width=2.5cm, height=2.5cm]{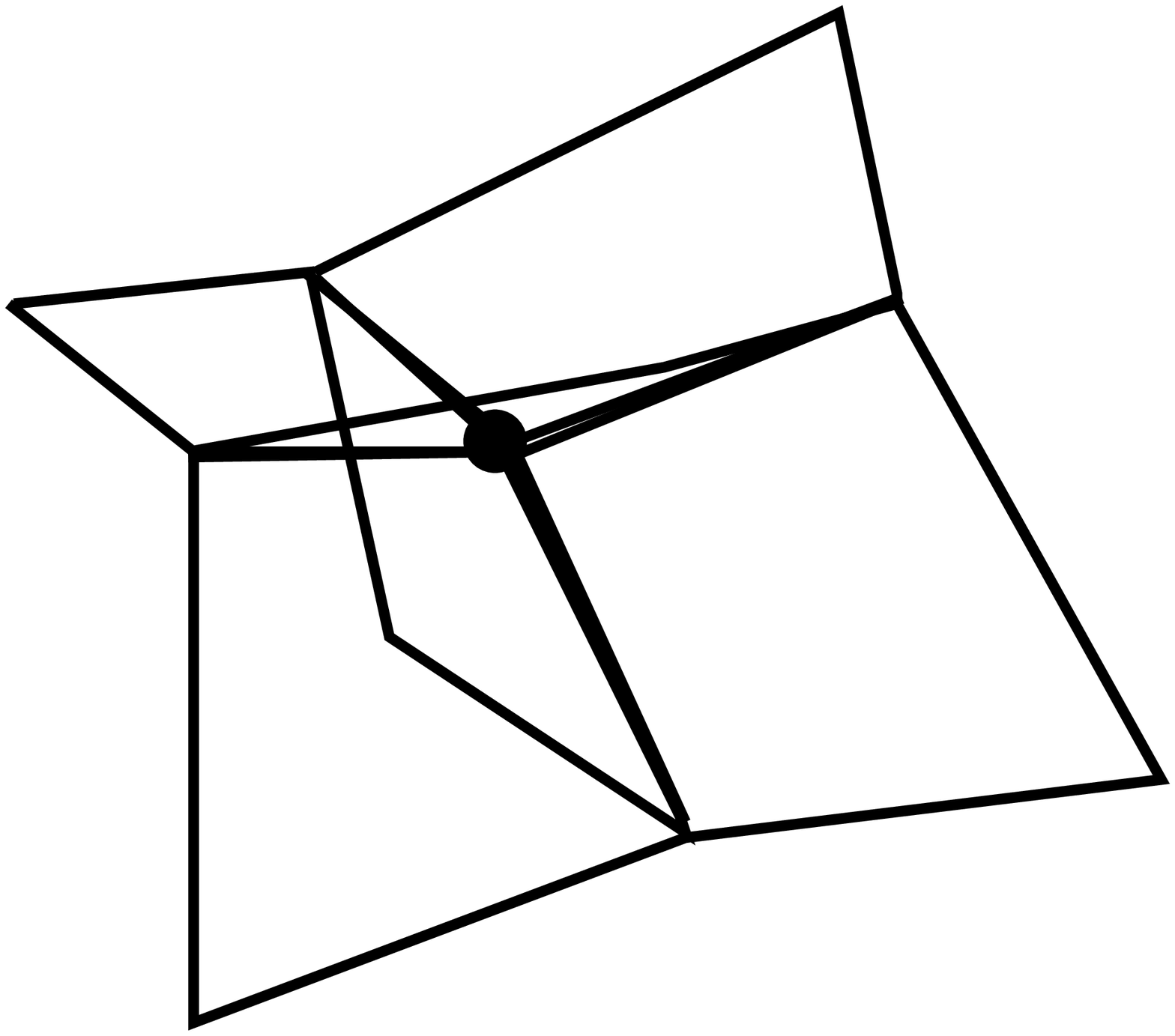}
\caption{Dual 2-complex}
\end{minipage}
\end{figure}

\vspace{-0.2cm}
Let us now identify the quantum amplitudes that the theory assigns to
the Feynman graphs. In configuration space the amplitude for
each 2-complex is:
$$
Z(\Gamma)\,=\, \left(\prod_{e\in \Gamma} \int d g_{e}\right)
\,\prod_{f}\,\delta (\prod_{e\in\partial f} g_{e} )
$$
which has the form of a lattige gauge theory partition function with
simple delta function weights for each plaquette (face of the
2-complex) and one connection variables for each edge; the delta
functions constraint the curvature on any face to be zero, as we
expect from 3d quantum gravity \cite{laurentPRI}.
To have the corresponding expression in momentum space, one expands
the field in modes $\phi(g_1,g_2,g_3)\,=\,\sum_{j_1,j_2,j_3}
\phi^{j_1j_2j_3}_{m_1n_1  m_2n_2
  m_3n_3}\,D^{j_1}_{m_1n_1}(g_1)D^{j_2}_{m_2n_2}(g_2)D^{j_3}_{m_3n_3}(g_3)$, 
where the $j$'s are irreps of $SU(2)$, obtaining, for the propagator, vertex and amplitude:

\begin{eqnarray*}
\mathcal{P} &=& \delta_{j_1\tilde{j}_1}\delta_{m_1\tilde{m}_1}
\delta_{j_2\tilde{j}_2} \delta_{m_2\tilde{m}_2}
\delta_{j_3\tilde{j}_3}\delta_{m_3\tilde{m}_3} \\ 
\mathcal{V} &=& \delta_{j_1\tilde{j}_1}\delta_{m_1\tilde{m}_1}
\delta_{j_2\tilde{j}_2} \delta_{m_2\tilde{m}_2}
\delta_{j_3\tilde{j}_3}\delta_{m_3\tilde{m}_3}\delta_{j_4\tilde{j}_4}\delta_{m_4\tilde{m}_4}\delta_{j_5\tilde{j}_5} \delta_{m_5\tilde{m}_5}
\delta_{j_6\tilde{j}_6}\delta_{m_6\tilde{m}_6} \left\{
\begin{array}{ccc} 
j_1 &j_2 &j_3
\\ j_4 &j_5 &j_6 
\end{array}\right\}
\\ Z(\Gamma)&=&\left(\prod_{f}\,\sum_{j_{f}}\right)\,\prod_{f}\Delta_{j_{f}}\prod_{v}\, \left\{ \begin{array}{ccc} 
j_1 &j_2 &j_3
\\ j_4 &j_5 &j_6 
\end{array}\right\} \end{eqnarray*}  
where $\Delta_j$ is the dimension of the representation $j$ and for
each vertex of the 2-complex we have a so-called $6j-symbol$ , i.e. a
scalar function of the 6 irreps meeting at that vertex. The
amplitude for each 2-complex is given then by a spin foam model, the
Ponzano-Regge model for 3d gravity without cosmological constant,
about which a lot more is known \cite{laurentPRI}.
The full theory is defined by the sum over all Feynman
graphs weighted by the above amplitudes:
$$ Z =\sum_{\Gamma}\,\frac{\lambda^N}{sym[\Gamma]}\,\left(\prod_{f}\,\sum_{j_{f}}\right)\,\prod_{f}\Delta_{j_{f}}\prod_{v}\,\left\{ \begin{array}{ccc} 
j_1 &j_2 &j_3
\\ j_4 &j_5 &j_6 
\end{array}\right\}_v. $$
This gives an un-ambigous realisation, in purely algebraic
and combinatorial terms, of the sum over both geometries and
topologies, i.e. of the third quantization idea. More precisely, it is
a simplicial third quantization, a {\it quantum
  field theory of simplicial geometry}, with fundamental classical
dynamical objects being triangles, quantum states given by collections
of quantum triangles represented as 3-valent spin networks, and
histories given by 3d triangulations.
\subsection{GFT: the general picture}

\begin{itemize}
\item GFT are thus a {\it local}, because one can easily consider bounded regions of space evolving or \lq
  timelike \rq boundaries, {\it discrete}, because it deals with
  discrete spacetimes, {\it algebraic and combinatorial}, because such are the variables in the theory, {\it 3rd quantization of gravity};
\item in fact, in GFTs both geometry and topology are dynamical, with precise quantum amplitudes assigned to each
  possible geometric and topological configuration of spacetime;
\item $D$-dimensional spacetime emerges via creation/annihilation of \lq\lq
  chunks\rq\rq of it, of spacetime quanta represented by (D-1)-simplices, as a Feynman diagram;
\item spacetime is therefore purely virtual in the quantum theory: just as the
  trajectory of a quantum particle or any specific interaction process
  in particle physics; no single spacetime
  configuration is realised as the \lq\lq truly existing\rq\rq
  spacetime, but all of them should be summed over to obtain a
  physical quantity, that is the probability of a specific bundary
  configuration;   
\item Quantum Gravity is described by an (almost) ordinary QFT,
  although with peculiar structure, and one that uses even a
  background metric ``spacetime'' (although here interpreted as an internal
  space only), given by a
  group manifold;
\item the GFT formalism has the potential to represent a unified framework for many
  current non-perturbative approaches to Quantum Gravity: Loop Quantum
  Gravity, Spin Foam models,
  Dynamical Triangulations, Quantum Regge Calculus, because its
  incorporates most of the basic ingredients on which these approaches
  are based, as one can easily realise: spin network states on the
  boundary, spin foam amplitudes for the histories, a dual sum over
  triangulations picture for its perturbative expansion, and a sum over
  geometric data, with amplitudes related to the Regge action for
  simplicial gravity.
\end{itemize}

\section{What lies ahead}
However fascinating the picture outlined above may be, we do not know
enough about group field theories to see it clearly in all its
detials, and therefore to fully believe. Even if lots is known
about the Feynman amplitudes of the theory, in various models in 3 and
4 dimensions\cite{review,alex,laurentgft}, in addition to what is known about matrix models in 2d,
it is probably fair to say that at present we do not know what a group
field theory is, and we can only deduce or guess some of its
properties on the basis of its Feynman amplitudes. In particular the
physical and geometric interpretation given above rests at present on
intuition only and it is not solidly based on mathematical results. 

We
list here a few of the directions that need to be explored if one has
to take GFTs seriously as a fundamental formulation of Quantum
Gravity.

First of all, we do not know much about the {\it classical} field
theories behind the perturbative expansion in spin foam we have
described: what are the solutions, in symmetric reduced cases at
least, of the classical equations of motion following from the above
actions? and what is their geometric interpretation? Work on this in
indeed in progress \cite{eteralaurent}. 

What
are the symmetries of the above action and the corresponding Ward
identities for Feynman graphs? Even in the simple 3d case it is not
easy to identify at the GFT level the translation symmetry that we
know it is present in the corresponding spin foam amplitudes
\cite{laurentPRI}. Most important, what is the GFT analogue of the
diffeomorphism symmetry of continuum gravity? what kind of other
symmetries should we expect in a theory in which topology change is
realised?

What is the physical meaning of the parameters of the action, i.e. in
the model we described, of the coupling cnstant $\lambda$? It can be
related to the cosmological constant in a simplicial gravity setting
\cite{DP-P} and/or it has the interpretation of a parameter governing
the strength of topology changing processes \cite{laurentgft}, but
much remains to be understood. 

At the quantum level, even though the
picture of spacetime as a process of creation/annihilation of
fundamental simplicial building blocks is appealing, it is at present
only a tentative picture; in fact, the Fock structure of the theory
has not been analysed in detail and rigorously, with a suitable
definition of creation/annihilation operators, on the basis of a
classical symplectic structure, and the definition of a
3rd quantized Fock vacuum. 

The relation with a canonical theory based
on spin network states is also unclear; while one can give a precise
and well-posed definition of a canonical inner product between
canonical states using a GFT \cite{laurentgft}, it would be
interesting to be able to extract from this the corresponding hamiltonian
constraint operator and compare it to the existing proposals
in loop quantum gravity; also, it would be interesting to compute the
corrections to the hamiltonian constraint equation coming from topology
changing terms, as in the formal continuum setting. 

There is much more
in a quantum field theory than its Feynman amplitudes in perturbative
expansion, and all this has still to be unveiled for the GFT case; in
particular, it is crucial for the issue of the continuum approximation
of these quantum gravity models to develop non-perturbative techniques,
probably after a suitable re-phrasing of them in statistical
mechanical terms, that would allow to study the phase structure of the
theory, and the emergence of a smooth spacetime in it, with continuum
General Relativity as an effective description of the degrees of
freedom of the theory in this phase. 

The coupling of matter and gauge
fields at the group field theory level, and the unification of these with gravity, is a whole area for
future developments, and work on this has just started \cite{kirill,
  us}. 

Also, as in ordinary quantum field theory, it should be
possible to define different types of transition amplitudes for the
same group field theories, with different uses and interpretation, as
seems to be confirmed by recent work \cite{io}. 

Finally, it remains to
be checked if the group field theory approach can maintain its promise of
being a general framework for as different approaches to quantum
gravity as loop quantum gravity, spin foam models, dynamical
triangulations and Regge calculus; many details of the links with
them have still to be understood and many gaps filled, but recent
results give reasons to hope \cite{io}.

\end{document}